\def \SAIT #1 #2 {{\em Mem.\ Soc.\ Astron.\ It.\/} {\bf #1}, #2}
\def \MESS #1 #2 {{\em The Messenger\/} {\bf #1}, #2}
\def \ASTRNACH #1 #2 {{\em Astron. Nach.\/} {\bf #1}, #2}
\def \AAP #1 #2 {{\em Astron. Astrophys.\/} {\bf #1}, #2}
\def \AAL #1 #2 {{\em Astron. Astrophys. Lett.\/} {\bf #1}, L#2}
\def \AAR #1 #2 {{\em Astron. Astrophys. Rev.\/} {\bf #1}, #2}
\def \AAS #1 #2 {{\em Astron. Astrophys. Suppl. Ser.\/} {\bf #1}, #2}
\def \AJ #1 #2 {{\em Astron. J.\/} {\bf #1}, #2}
\def \ANNREV #1 #2 {{\em Ann. Rev. Astron. Astrophys.\/} {\bf #1}, #2}
\def \APJ #1 #2 {{\em Astrophys. J.\/} {\bf #1}, #2}
\def \APJL #1 #2 {{\em Astrophys. J. Lett.\/} {\bf #1}, L#2}
\def \APJS #1 #2 {{\em Astrophys. J. Suppl.\/} {\bf #1}, #2}
\def \APSS #1 #2 {{\em Astrophys. Space Sci.\/} {\bf #1}, #2}
\def \ASR #1 #2 {{\em Adv. Space Res.\/} {\bf #1}, #2}
\def \BAIC #1 #2 {{\em Bull. Astron. Inst. Czechosl.\/} {\bf #1}, #2}
\def \JSQRT #1 #2 {{\em J. Quant. Spectrosc. Radiat. Transfer\/} {\bf #1}, #2}
\def \MN #1 #2 {{\em Mon. Not. R. Astr. Soc.\/} {\bf #1}, #2}
\def \MEM #1 #2 {{\em Mem. R. Astr. Soc.\/} {\bf #1}, #2}
\def \PLR #1 #2 {{\em Phys. Lett. Rev.\/} {\bf #1}, #2}
\def \PASJ #1 #2 {{\em Publ. Astron. Soc. Japan\/} {\bf #1}, #2}
\def \PASP #1 #2 {{\em Publ. Astr. Soc. Pacific\/} {\bf #1}, #2}
\def \NAT #1 #2 {{\em Nature\/} {\bf #1}, #2}

\documentstyle{memsait}
\input epsf.sty
\begin{opening}
\title{THE EVOLUTION OF TYPE 1 AGN AT 15 $\mu$m FROM THE ELAIS} 
\author{F. La Franca$^1$, I. Matute$^1$, C. Gruppioni$^2$,
D. Alexander$^3$, {\it + ELAIS consortium} }
\institute{$^1$Dipartimento di Fisica, Universit\`a Roma Tre, Via della Vasca Navale 84, 00146, Roma, Italy\\
$^2$Osservatorio di Padova, Vicolo dell'Osservatorio 35, 35122, Padova, Italy\\
$^3$SISSA, Via Beirut 4, 00000, Trieste, Italy}
\date{} 
\end{opening}

\begin{document}

\oddpagefooter{\sf Mem. S.A.It.}{}{\thepage}
\evenpagefooter{\thepage}{}{\sf Mem. S.A.It.}
\ 
\bigskip

\begin{abstract}

We present the first preliminary estimate of the evolution of type 1
AGNs at 15 $\mu$m. A new sample of sources selected from the ELAIS-S1
survey has been used together with the RMS sample. The ELAIS-S1 survey
covers an {\it effective} area of 2.2 deg$^2$ (total: 4 deg$^2$) down
to 1.2 mJy at 15 $\mu$m. Spectroscopic identifications have been
carried out in the range 17.2$<$R$<$20.0 at the 3.6m/ESO, NTT/ESO and
2dF/AAT. The luminosity function is steeper at bright luminosities in
comparison to the previous estimates of Rush, Malkan
\& Spinoglio (1993). The evolution is compatible with a Pure
Luminosity Evolution model with $L(z)\propto L(0)(1+z)^{3.4}$.
\end{abstract}

\section{The European Large Area ISO Survey (ELAIS)}

ELAIS is a collaboration involving 25 European Institutes, led from
Imperial College. ELAIS is one of the largest single Open Time project
conduced by ISO, mapping an area of 12 squares degrees at 15$\mu $m
with ISOCAM and at 90$\mu $m with ISO-PHOT. Secondary surveys in other
ISO bands were undertaken by the ELAIS team within the fields of the
primary survey, with 6 degrees being covered at 6.7$\mu $m and 1
degree at 175$\mu $m (Oliver et al, 2000).  Four main fields were
chosen (N1,N2,N3 in the north hemisphere and S1
 in the south) at
high Ecliptic latitudes
($|\beta|>40^{\scriptscriptstyle{o}}$).

The work here presented is the result of the follow-up campaign
carried out in the field S1 (J2000, $\alpha:
00^{\scriptscriptstyle{h}}34^
{\scriptscriptstyle{m}}44^{\scriptscriptstyle{s}} , \delta:
{-43}^{\scriptscriptstyle{O}} 34^{\scriptscriptstyle{'}}44''$).  An
initial catalog was produced by the Imperial College using their data
reduction technique (``Preliminary Analysis''). Optical
identifications were possible
 thanks to an extensive R-band CCD
survey, performed with the
 ESO/Danish 1.5m telescope.  The
spectroscopic follow-up program resulted in $\sim$130 redshift
identifications with the 2dF/AAT, 3.6/ESO and NTT/ESO.  The sample is
flux limited at 15$\mu$m down to 1.2 mJy, and in the R-band to
magnitudes between 17.2 and 20.0.

\section{The luminosity function at 15 $\mu$m}

In order to compute the luminosity function (LF) of type 1 AGNs (AGN1)
at 15 $\mu$m, we added to our sample the AGN1 from the catalog of
Rush, Malkan \& Spinoglio (1993, RMS) as representative of the local
population. The RMS consists of a sample of galaxies selected a 12
$\mu $m from the IRAS PSCv2 complete down to 0.3 Jy.

We characterize our sources in the Infrared with a single SED.
Ground-based observations (Roche et al. 1991; Moorwood 1986) and the
latest ISO observations on Type 1 AGNs find that the Mid-Infrared
(MIR) broadband emission features (from PAH molecules) are generally
absent in this kind of sources (Lutz et al. 1997a; Genzel et
al. 1998). For this reason the mean SED for radio-quiet Quasars from
Elvis et al. (1994) was chosen.  The standard K-correction (in
magnitudes) was computed following Lang (1980).  In order to compute
the 15$\mu$m luminosities and $K$-corrections, we convolved the SED
from Elvis et al. (1994) with the IRAS 12$\mu$m and ISOCAM 15$\mu$m
bandpasses.

We assumed a simple double power-law luminosity function (LF) and
pure luminosity evolution (PLE) of the type:
\vspace{0.1cm}
\begin{eqnarray*}
\Phi(\mathrm{L}) & = & \mathrm{C} \, \mathrm{L}_{*}^{(\alpha - 
\beta)} \, \mathrm{L}^{-\alpha} \hspace{0.7cm} \mathrm{for} \hspace{0.7cm} 
\mathrm{L} \leq \mathrm{L}_{*}, \\
\Phi(\mathrm{L}) & = & \mathrm{C} \, \mathrm{L}^{- \beta} 
 \hspace{1.8cm} \mathrm{for} \hspace{0.7cm} \mathrm{L}> \mathrm{L}_{*} 
\end{eqnarray*}
\vspace{0.1cm}
where $\alpha$ and $\beta$ are the slopes of the faint and 
bright end of the LF, L$_*$ the luminosity of the 'knee' of the LF and C the normalization
factor. The evolution in luminosity is given by the relation:

\begin{center}
$\mathrm{L}_{*}(z) = \mathrm{L}_{*}(0)\, (1+z)^{\kappa}.$
\end{center}

A parametric, unbinned maximum likelihood method was used to fit the
evolution and luminosity function parameters simultaneously (Marshall
et al. 1983) at 15$\mu$m.  Since ELAIS identifications were not only
flux limited at 15$\mu$m, but also in their R-band magnitude (17.2 $<$
R $<$ 20.0), a factor $\mathbf{\Theta}$ was introduced in the function
'S' to be minimized

\vspace{0.2cm
\begin{center}}
\begin{displaymath}
\mathrm{S}=-2 \,\sum_{i=1}^{N} ln[\Phi(z_{i},L_{i})] + \int\!\!\!\int 
\Phi(z,L) \Omega(z,L) \, \mathbf{\Theta} \mathrm{(z,L) \, \frac{\textstyle dV}
{\textstyle dz}dzdL}
\end{displaymath}
\end{center}
\vspace{0.2cm}
and only applied to the ELAIS sample.
This factor represents the conditional probability of a source with a
given 15 $\mu$m luminosity to have R magnitude in the range (17.2,20.0):

\begin{center}
$\mathbf{\Theta}\mathrm{  \, (17.2<R<20.0 \, \mid \mathrm{L}_{15})\,} ,$
\end{center}
computed using the local (RMS) L$\mathrm{_R}$/L$_{15}$ distribution, 
supposed constant with z.

\section{Results and comments}

For the first time, the evolution of the Type 1 AGNs luminosity
function at 15$\mu$m is derived from a list of sources selected in the
Infrared (Fig.1).  The LF evolves according to a PLE model with
$\kappa \simeq 3.4$, an evolution similar to the one derived in the
optical domain (e.g. La Franca \& Cristiani, 1997).

The slope of the previous derivation of the LF from RMS has a flatter
slope in the bright part ($\beta = 2.1$) in comparison with our
determination ($\beta = 2.5$). This overestimate of the density of
bright AGN1 by RMS is due to the effects of evolution inside the
redshift bins used with the 1/V$_{a}$ method.


\begin{figure}
\epsfysize=6cm 
\hspace{3.0cm}
\epsfbox{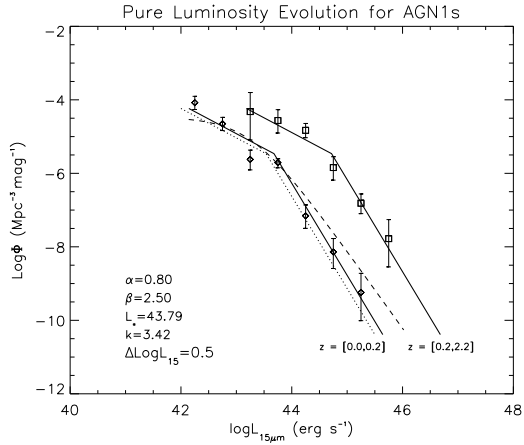} 
\caption[h]
{Best-fitting PLE model for 2 redshifts intervals (\textit{solid
lines}).The points (\textit{rhombuses, squares}) correspond to the
space-densities of the observed sources, corrected for evolution
within the redshift intervals.  As a comparison, the RMS fit of the
luminosity function (\textit{dashed line}) and our LF at z=0
(\textit{dots}) are also plotted.}
\end{figure}



\end{document}